\documentclass[10pt]{article}

\usepackage[T1]{fontenc}
\usepackage[utf8]{inputenc}

\usepackage{amsmath}
\usepackage{amssymb}

\usepackage{algorithm2e}
\makeatletter
\let\listofalgorithms\@undefined
\makeatother
\usepackage{algorithm}

\usepackage[skip=0pt]{caption}
\usepackage[aboveskip=0pt,belowskip=6pt]{subcaption}
\usepackage[margin=1.5in]{geometry}

\usepackage{cancel}

\input{mathdefs}

\auxfun{RTM}
\auxfun{prepare}
\auxfun{effect}
\auxfun{eval}
\auxfun{apply}
\auxfun{merge}
\auxfun{inc}
\auxfun{dec}
\auxfun{fetch}
\auxfun{rd}
\auxfun{elems}
\auxfun{op}
\auxfun{subredundant}
\auxfun{substable}
\auxfun{subeval}
\auxfun{elements}
\auxfun{size}
\auxfun{add}
\auxfun{enable}
\auxfun{disable}
\auxfun{upd}
\auxfun{set}
\auxfun{del}
\auxfun{low}
\auxfun{rmv}
\auxfun{remove}
\auxfun{clear}
\auxfun{cbcast}
\auxfun{tcbcast}
\auxfun{tcdeliver}
\auxfun{cdeliver}
\auxfun{tcstable}
\auxfun{stabilize}

\auxfun{rcbBcast}
\auxfun{rcbDeliver}
\auxfun{ercbBcast}
\auxfun{ercbDeliver}
\auxfun{ercbStable}
\auxfun{preprocess}
\auxfun{process}
\auxfun{postprocess}

\auxfun{stable}
\auxfun{deliv}
\auxfun{src}
\auxfun{redundant}
\auxfun{operation}

\newcommand\R{\mathrel{\mathsf{R}}}

\newcommand\polog{PO-Log\xspace}

\newcommand\gctr{GCounter\xspace}
\newcommand\pnctr{PNCounter\xspace}
\newcommand\gset{GSet\xspace}
\newcommand\twopset{2PSet\xspace}

\newcommand\ewfl{EWFlag\xspace}
\newcommand\dwfl{DWFlag\xspace}
\newcommand\mvreg{MVRegister\xspace}
\newcommand\awset{AWSet\xspace}
\newcommand\rwset{RWSet\xspace}

\newtheorem{definition}{Definition}[section]

\begin{document}
\title{Pure Operation-Based Replicated Data Types\footnote{
  The work presented was partially supported by EU FP7 SyncFree 
project (609551), EU H2020 LightKone project (732505), and SMILES 
line in project TEC4Growth (NORTE-01-0145-FEDER-000020.}}

\author{Carlos~Baquero,~Paulo~S\'{e}rgio~Almeida~and~Ali~Shoker\\
  HASLab, INESC TEC and Universidade do Minho,\\ Braga, Portugal.  }

\maketitle

\begin{abstract}

Distributed systems designed to serve clients across the world often make use
of geo-replication to attain low latency and high availability.  Conflict-free
Replicated Data Types (CRDTs) allow the design of predictable multi-master
replication and support eventual consistency of replicas that are allowed to
transiently diverge. CRDTs come in two flavors: state-based, where a state is
changed locally and shipped and merged into other replicas; operation-based,
where operations are issued locally and reliably causal broadcast to
all other replicas. 
However, the standard definition of op-based CRDTs is very encompassing,
allowing even sending the full-state, and thus imposing storage and
dissemination overheads as well as blurring the distinction from state-based
CRDTs.
We introduce pure op-based CRDTs, that can only send operations to other
replicas, drawing a clear distinction from state-based ones.  Data types with
commutative operations can be trivially implemented as pure op-based CRDTs
using standard reliable causal delivery; whereas data types having
non-commutative operations are implemented using a \emph{\polog}, a partially
ordered log of operations, and making use of an extended API, i.e., a \emph{Tagged
Causal Stable  Broadcast (TCSB)}, that provides extra causality information
  upon delivery and later informs when delivered messages become causally
  stable, allowing further \emph{\polog} compaction. 
The framework is illustrated by a catalog of pure op-based specifications
for classic CRDTs, including counters, multi-value registers, add-wins and remove-wins sets.

\end{abstract}

\section{Introduction}

Eventual consistency~\cite{rep:syn:pan:1624} is a relaxed consistency model that
is often adopted by large-scale distributed
systems~\cite{riak+crdt,Bailis:2013:ECT:2460276.2462076,syn:optim:rep:1433,app:rep:optim:1606}
where losing availability is normally not an option, whereas delayed consistency
is acceptable. In eventually consistent systems, data replicas are allowed to
temporarily diverge, provided that they can eventually be reconciled into a
common consistent state.  Reconciliation (or merging) used to be error-prone,
being application-dependent, until new data-type-dependent models like the
Conflict-free Replicated Data Types~(CRDTs)~\cite{rep:syn:sh138,syn:rep:sh143}
were introduced. CRDTs allow both researchers and practitioners
to design correct replicated data types that are always available, and are
guaranteed to eventually converge once all operations are known at replicas.
The concept of CRDT has been deployed in practice by industry, initially in Key-Value Stores~\cite{riak+crdt}, and have since been ported to multiple languages and production platforms. 

CRDTs support two complementary designs: operation-based (or simply, op-based)
and state-based. As the name suggests, the former are based on dissemination of operations and the later on shipping state that results from locally applied operations. 
In op-based
designs~\cite{alg:rep:sh132,syn:rep:sh143}, the execution of an operation is
done in two phases: \emph{prepare} and \emph{effect}. The former is
performed only on the local replica and looks at the operation and current
state to produce a message that aims to represent the operation, which is then
shipped to all replicas.  Once received, the representation of the
operation is applied remotely using \emph{effect}. Different replicas are
guaranteed to converge as long as messages are disseminated through a
reliable causal broadcast messaging
middleware~\cite{pro:pan:160}, and \emph{effect} is designed to be commutative for concurrent
operations. On the other hand, in a state-based
design~\cite{app:1639,syn:rep:sh143}, an operation is only executed on the
local replica state. A replica propagates its local changes to other replicas
through shipping its entire state. A received state is incorporated with the
local state via a \emph{merge} function that, deterministically, reconciles
the \emph{merged} states. To maintain convergence, \emph{merge} is defined as
a join: a least upper bound over a
join-semi-lattice~\cite{app:1639,syn:rep:sh143}.
Typically, state-based CRDTs support ad hoc dissemination of states and can
handle duplicate and out-of-order delivery of messages without breaking causal
consistency; however, they impose complex state designs and store extra
meta-data. 
On the other hand, in the systems where the message dissemination
layer guarantees reliable causal broadcast, operation-based CRDTs have more
advantages as they can allow for simpler implementations, concise replica
state, and smaller messages.

In standard op-based CRDTs the designer is given much freedom in defining
\emph{prepare}, namely using the state in an arbitrary way.  This is needed to
have the \emph{effect}s of concurrently invoked data-type operations commute,
and thus provide replica convergence despite the absence of causality
information in current causal delivery APIs. This forces current op-based
designs to include causality information in the state to be used in
\emph{prepare}, sent in messages, and subsequently used in \emph{effect}. The
designer ends up intervening in many components (the state, \emph{prepare},
\emph{effect}, and \emph{query} functions) in an ad hoc way.  This can result
in large complex state structures and also large messages.

Currently, a \emph{prepare} not only builds messages that duplicate the
information already present in the middleware (even if it is not currently
made available), but causality meta-data is often incorporated in the object
state, hence, reusing design choices similar to those used in state-based
approaches. Such designs impose larger state size and do not fully exploit causal delivery information.
This freedom in current op-based designs is against the spirit of `sending
operations', and leads to confusion with the state-based approach. Indeed, in
the current op-based framework, a \emph{prepare} can return the full state,
and an \emph{effect} can do a full state-merge (which mimics a state-based
CRDT)~\cite{app:1639,syn:rep:sh143}.

We believe that the above weaknesses can be avoided if the causality meta-data
can be provided by the messaging middleware. Causal broadcast implementations
already possess that information internally, but it is not exposed to clients.
In this paper we propose and exploit such an extended API to achieve both
simplicity and efficiency in defining op-based CRDTs.

We introduce a \emph{Pure} Op-Based CRDT framework, in which \emph{prepare}
cannot inspect the state, being limited to returning the operation (including
potential parameters). The entire logic of executing the operation in each
replica is delegated to \emph{effect}, which is also made generic (i.e., not
data type dependent).  For pure op-based CRDTs, we propose that the object
state is a \emph{partially ordered log of operations -- a \polog}.  Causality
information is provided by an extended messaging API: \emph{Tagged
Causal Stable Broadcast} (TCSB). We use this information to preserve
convergence and also design compact and efficient CRDTs through a
\emph{semantically based \polog compaction} framework, which makes use of a
data type-specific \emph{obsolescence} relation, defined over
timestamp-operation pairs.

Furthermore, we propose an extension that improves the design and implementation of
op-based CRDTs through decomposing the state into two components: a \polog (as
before), and a causality-stripped-component which, in many cases, will be
simply a standard sequential data type.
For this, we introduce the notion of \emph{causal stability}, to be provided
by the TCSB middleware.
The idea is that operations are kept only transiently in the \polog, but once
they become causally stable, causality meta-data is stripped, and the
operations are stored in the sequential data type. This reduces the
storage overhead to a level that was never achieved before in CRDTs, neither
state-based nor op-based.

\section{System Model and Notations}
\label{sec:model}

The system is composed of a fixed set of nodes, each is associated with a 
globally unique identifier in a set $I$. Nodes execute operations at different 
speeds and communicate using asynchronous message passing.
Messages can be lost, reordered, or duplicated, and the system can experience
arbitrary, but transient, network partitions. To ensure exactly-once delivery, 
message sending is abstracted by a
reliable causal broadcast~\cite{birman:1987:reliable,golding1992weak}.
A node can fail by crashing and can recover later on. Upon recovery, the last
durable state of a node is assumed to be intact (not destroyed). We do not
consider Byzantine or malicious faults.

For presentation purposes, and without loss of generality, we consider a single
data type instance that is fully replicated at each node. All replicas (of a data type) are 
initially equivalent. Once a data type operation is locally applied on a replica, 
the latter can diverge from other replicas, but replicas may 
eventually convergence as
operations arrive everywhere. The local application of an operation is atomic
on each replica (i.e., upon a crash, no state `in the middle of an operation'
can be seen in the durable state).

\subsection{Definitions and Notations}
$\Sigma$ denotes the type of the state. $\mathcal{P}(V)$ denotes a \emph{power
set} (the set of all subsets of $V$), where $V$ is a set of any type.
The initial state of a replica $i$ is denoted by $\sigma^0_i\in \Sigma$.
Operations belong to a set $O$ and can include arguments, in which case they
are surrounded by square brackets, e.g., $\inc$ and $[\add,v]$.
On the other hand, the notation $o[j]$ refers to the $j^{th}$
element in a list that comprises the operation name and subsequent arguments (analogous to
the $argv[]$ used in several programming languages); in particular, $o[0]$ 
refers to the operation name.
We use total functions $K \to V$ and maps (partial functions) $K \map V$
from keys to values, both represented as sets of pairs $(k, v$). Given a
function $m$, the notation $m\{k \mapsto v\}$ maps a specific key $k$
to $v$, and behaves like $m$ on other keys.

\section{Background}
\label{sec:back}

By tracking divergence, conflict-free replicated data-types allow local operation even when no communication is possible. We have seen that CRDTs can be approached by state-based\footnote{For more information
	about state-based CRDTs and the optimized variant Delta-State CRDTs,
	the reader may refer to ~\cite{rep:syn:sh138,syn:rep:sh143} 
	and~\cite{delta-states:Almeida:2016,deltas:Almeida:2015}.} or op-based frameworks, and both approaches share the same high-availability properties and resiliency to partitions. Onwards, this paper will focus on advancing the definition and possibilities of the operation-based approach.  

\subsection{Operation-Based CRDTs}
In op-based CRDTs~\cite{rep:syn:sh138,syn:rep:sh143}, each replica
maintains a local state, and is subject to clients \emph{query} and \emph{update}
operations that are executed locally as soon as they arrive. 
In particular, update operations that are received through clients are 
disseminated to other replicas, via a reliable causal broadcast middleware, 
in a form of operation name, arguments (if exist), and possibly some
meta-data~\cite{rep:syn:sh138,syn:rep:sh143}.
Once all replicas receive and execute all update operations, they
eventually converge to a single state  as long as concurrent update operations commute, i.e. produce the same effect if delivered in different orders. 
Operation dissemination must be reliable, and most data-types require broadcast delivery to respect causal order. This also ensures that resulting states are causally consistent.

\begin{algorithm}[H]
\DontPrintSemicolon
\SetKwBlock{state}{state:}{}
\SetKwBlock{on}{on}{}
\state{
  $\sigma_i \in \Sigma$ \;
}
\on({$\operation_i(o)$:}){
	$m:=\prepare_i(o, \sigma_i)$ \;\\
	$\cbcast_i(m)$ \;
}
\on({$\cdeliver_i(m)$:}){
  $\sigma_i := \effect_i(m, \sigma_i)$ \;
}
\BlankLine
\caption{Distributed algorithm at node $i$ showing the interplay between a standard
reliable causal broadcast API ($\cbcast$ and $\cdeliver$) and op-based CRDTs 
($\prepare$ and $\effect$).}
\label{algo:op-rcb}
\end{algorithm}

\begin{figure}[h]
	\centering
\[{\centering
	\begin{array}{cl}
		\Sigma&:\text{\small State type, $\sigma_i$ is an instance}\\ \prepare_i(o,\sigma_i)&:\text{\small Prepares a message $m$  given an operation $o$}\\
		\effect_i(m,\sigma_i)&:\text{\small Applies a \emph{prepared} message $m$ on a state}\\
		\eval_i(q,\sigma_i)&:\text{\small Read-only evaluation of query $q$ on a state}
	\end{array}}
\]
	\caption{The general scheme of an op-based CRDT.}
	\label{fig:op-crdt}
\end{figure}

More specifically, the typical scheme of op-based CRDT is depicted 
in Figure~\ref{fig:op-crdt}, while Algorithm~\ref{algo:op-rcb} depicts the
interplay between the reliable causal broadcast middleware and the CRDT.
When an update operation $o$ is issued at some node $i$ having state $\sigma_i$, 
the function $\prepare_i(o,\sigma_i)$ is called returning a message $m$. 
This message $m$ is then broadcast by calling $\cbcast_i(m)$ that is provided
by the reliable causal broadcast API. 
Once $m$ is delivered via $\cdeliver_j(m)$ at each destination
node $j$ (including $i$ itself), $\effect_j(m,\sigma_j)$ is called, returning
a new replica state $\sigma'_j$. For each node that broadcasts a given
operation, the broadcast event, the corresponding local delivery, and the
$\effect$ on the local state are executed atomically. 
When a query operation $q$ is issued, $\eval_i(q, \sigma_i)$ is invoked, and no
corresponding broadcast occurs. $\eval$ takes the query and the state as input 
and returns a result, leaving the state unchanged. 
We further explain this through the following simple example.

\emph{Example on Op-Based Counters:}
Figure~\ref{fig:op-ctr} represents an op-based increment-only counter. The
state represents an integer that is initialized to $0$. The $\prepare$ function
only returns $m=\inc$ for an invoked $\inc$ operation by a client. 
The middleware disseminates $m$ to all replicas, whereas, $\effect$ increments 
the counter state at each replica. 
The query operation $\af{value}$ simply returns the value of the counter.\\

\subsection{The case for a new op-based CRDT model}
The above example on counters is useful to explain the op-based CRDT model, 
but it falsely gives
the impression that the CRDT design is generally straightforward and lightweight.
Unfortunately, this is not the case for more complex data types where operations
are not commutative, e.g., in sets, maps, etc. 
For instance, consider the \emph{Add-Wins Set} design shown in Figure~\ref{fig:op-set} 
and having the following semantics: an $\add$ operation 
prevails if it is concurrent with a $\remove$ (an alternative remove-wins design is shown later). 
Since $\add$ and $\remove$ are not commutative operations, the causal order of 
their execution must be ensured to guarantee that different replicas converge. 
Capturing this information leads to a more complex 
design through adding 
extra meta-data to the data type state (see $\Sigma$) as well as to the exchanged messages 
(see the output of $\prepare$). 

Indeed, after analyzing several op-based data types introduced 
in~\cite{rep:syn:sh138,syn:rep:sh143}, we observed three general trends (we discuss
these in details in Section~\ref{sec:discuss}): 
first, the entire design is data-type-dependent since all the components (i.e.,
$\Sigma, \prepare, \effect,$ and $\eval$) can change per data type; 
second, designing (or understanding) an efficient data type becomes cumbersome;
and third, this design induces a substantial communication overhead since 
$\prepare$ returns additional meta-data to be sent (along with the operation 
name and parameters). The pure op-based CRDT model that we introduce in this
paper tries to address these challenges.

\begin{figure}[t]
\centering
\begin{subfigure}[b]{.35\textwidth}
\centering
\begin{eqnarray*}
\Sigma = \nat && \sigma_i^0 = 0 \\
\prepare_i(\inc,n ) & = & \inc \\
\effect_i(\inc, n ) & = & n+1 \\
  \eval_i(\af{value},n ) & = & n\\
\end{eqnarray*}
\caption{Op-based increment-only counter.}
\label{fig:op-ctr}
\end{subfigure}\\
\begin{subfigure}[b]{.60\textwidth}
\centering
\begin{eqnarray*}
\Sigma = \nat \times \pow{I \times \nat \times V} &&
\sigma_i^0 = (0,\{\})\\
\prepare_i([\add,v],(n,s)) & = & [\add,v,i,n+1] \\
\effect_i([\add,v,i',n'], (n,s)) & = & (n' \kw{if} i = i' 
\kw{otherwise} n\\&&, s \cup \{(v,i',n')\} )\\
\prepare_i[\remove,v],(n,s)) & = & [\remove,\{ (v',i',n') \in s | v'= v\}]\\
\effect_i([\remove,r],  (n,s )) & = & (n, s \setminus r ) \\
\eval_i(\elems,(n,s )) & = & \{ v | (v,i',n') \in s\}\\
\end{eqnarray*}
\caption{Op-based Add-Wins Set (embeding causality)}
\label{fig:op-set}
\end{subfigure}\\
\caption{An example of two standard op-based CRDTs showing their 
	design complexity (when causality is needed) and their
	data-type-dependent nature.}
\label{fig:op-based}
\end{figure}

\section{Pure Op-based CRDTs}
\label{sec:pure}

In this section, we introduce \emph{pure op-based CRDTs}. We show that
data types whose operations are commutative are natively ``pure''; whereas
for other data types to be pure, they need to leverage some information 
channeled through the dissemination middleware.

\begin{definition}[Pure op-based CRDT]
An op-based CRDT is \emph{pure} if disseminated messages contain only 
the operation and its potential arguments. 
\end{definition}

Therefore, considering the general op-based CRDT scheme introduced 
in Figure~\ref{fig:op-crdt}, and given an operation $o$ and a state $\sigma$,
$\prepare$ is always defined as:
\[
\prepare(o, \sigma) = o
\]

In addition to the generic nature of this property across different data types,
at least two other benefits can be identified. 
First, $\prepare$ cannot build an arbitrary 
message depending on the current state which avoids introducing design or 
implementation bugs. Second, the operation can immediately
be broadcast without even reading the replica state $\sigma$. 
(Indeed, the parameter $\sigma$ of $\prepare(o, \sigma)$ can be omitted in
pure op-based CRDTs, but we keep it for consistency and clarity.). This has
good performance implications on the sender side as the access to persistent 
storage and serialization/deserialization are avoided. 

Back to the data type examples described in Figure~\ref{fig:op-based}, we notice
that the counter data type is pure op-based since $\prepare=\inc$.
On the contrary, the Add-Wins Set in Figure~\ref{fig:op-set} is not pure 
since a $\prepare$ builds a set of triples present in the current state, 
to be used by other replicas when performing $\effect$ (either for $\add$ 
or for $\rmv$). To understand this difference, we make a distinction between
two categories: commutative and non-commutative data types.

\begin{definition}[Commutative data type]
A concurrent data type is commutative iff: 
\begin{enumerate}
	\item for any operations $f$ and $g$, their sequential 
		invocation commutes: $f(g(\sigma))=g(f(\sigma))$, and
	\item all concurrent invocations are defined as equivalent to some linearization.
\end{enumerate}
\end{definition}

While the first property removes the ordering overhead of operations (as no 
ordering assumptions are required), the second property makes the pure op-based 
design of a commutative data type straightforward as we explain next.

\subsection{Pure designs for commutative data types}
Commutative data types can natively be designed as pure op-based CRDTs on top
of a standard reliable causal broadcast~\cite{causal-multicast:Birman:1991}.
The reason is that commutative data types reflect a \emph{principle of permutation
equivalence}~\cite{rep:opt:sh150}: if all sequential
permutations of updates lead to equivalent states, then it should also hold
that concurrent executions of the updates lead to equivalent states.
As the extension to concurrent scenarios follows directly from their
sequential definition (with no room for design choices), commutative data types
can have a standard sequential specification and design. Consequently,
a pure op-based CRDT design becomes trivial: since a message returned from
$\prepare$, and containing the operation name and arguments, will arrive 
exactly once at each replica (through the reliable causal broadcast), 
it is enough to have $\effect$ only apply the received operation to the standard 
sequential data type state, i.e., defining for any operation $o$:
\[
\effect(o, \sigma) = o(\sigma)
\]

\emph{Examples on commutative data types:} Two examples are presented in
Figure~\ref{fig:pure-commutative}: a PN-Counter with $\inc$ and $\dec$
operations, and a Grow-only Set (G-Set) with $\add$ operation. Both cases use a
standard sequential data type for the replica state $\Sigma$, 
and applying $\effect$ is just invoking the corresponding standard 
operation ($+,-,\cup$, etc.) to 
the sequential data type. Both examples exploit commutativity 
and rely on the exactly-once delivery, leading to a trivial pure design. 

\begin{figure}[t]
\begin{subfigure}[b]{.47\textwidth}
\centering
\begin{eqnarray*}
\Sigma = \nat &&
\sigma_i^0 = 0 \\
\prepare_i(o, \sigma_i) & = & o \text{\quad($\inc$ or $\dec$)} \\
\effect_i(\inc, n ) & = & n+1 \\
\effect_i(\dec, n ) & = & n-1 \\
\eval_i(\af{value},n ) & = & n\\
\end{eqnarray*}
\caption{Pure PN-Counter}
\label{fig:oppnctr}
\end{subfigure}
\begin{subfigure}[b]{.47\textwidth}
\centering
\begin{eqnarray*}
\Sigma = \pow{V} &&
\sigma_i^0 = \{\} \\
\prepare_i(o, \sigma_i) & = & [\add, v] \\
\effect_i([\add, v], s) & = & s \union \{v\} \\
\eval_i(\elems, s) & = & s\\
\end{eqnarray*}
\caption{Pure Grow-Only Set}
\label{fig:opgset}
\end{subfigure}
\caption{Pure op-based CRDTs for commutative data types.}
\label{fig:pure-commutative}
\end{figure}

\subsection{The challenge of non-commutative data types}
\label{sec:challenges}
In the case where the operations of a data type are not commutative, 
e.g., $\add$ and $\rmv$ in a set ($\add(v, \rmv(v, s)) \neq \rmv(v, \add(v,
s))$), an $\effect$ cannot simply apply the operations over a sequential 
data type as done in the previous section: $o(\sigma)$. This is due to
two reasons, that we explain below, on which we base our extended model.

The first reason is that the messages corresponding to concurrent 
operations can be delivered in different orders on different 
replicas. Since the operations do not commute (even if the 
semantics of concurrent invocations can be defined as equivalent to some 
linearization of those operations), simply applying them in different 
orders on different replicas makes these replicas diverge.
Therefore, 
convergence is only guaranteed if $\effect$ is commutative, and therefore, cannot be
defined directly as operations that are not commutative themselves (e.g., simple
$\add$ and $\remove$ operations).
To address this point, $\effect$ must have partial-ordering knowledge about operations
and a corresponding logic in execution. While the latter suggests changing the
design of $\effect$, the former may require exchanging some ordering meta-data.
However, since such information
about the partial order is already present in the meta-data of the causal delivery
middleware, we propose to extend the API of the middleware (presented in 
Section~\ref{sec:ercb}), which we call 
\emph{Tagged Causal Stable Broadcast} (TCSB), to leverage this meta-data that can then be 
used in designing general non-commutative pure op-based CRDTs.

The other reason is that, indeed, there are useful concurrent data types in which
the outcomes of concurrent executions are not equivalent (on purpose) to some 
linearization. The best example is the Multi-Value Register (MVRegister) that is used in 
Amazon's Dynamo system to model the shopping cart~\cite{app:rep:optim:1606}. In MVRegister, invoking
a read operation following two concurrent writes in its causal past would return a 
set with \emph{both} values written. This behavior does not arise under a sequential
specification of a register, and the state must retain concurrent operations.
We address this challenge by proposing (in Section~\ref{sec:polog}) the type 
of a CRDT (i.e., $\Sigma$) 
to be a partially-ordered log of operations provided that $\effect$ understands 
how to execute these operations, thanks to the TCSB meta-data.

\section{Tagged Causal Stable Broadcast (TCSB)}
\label{sec:ercb}
Reliable Causal Broadcast (RCB) is a prominent dissemination abstraction that is
often used in AP systems (in the CAP~\cite{cap:Gilbert:2002} context) to 
ensure causal delivery of messages, being the strongest consistency 
model in always-available systems that eventually 
converges~\cite{syn:rep:1677,causal:Attiya:2017}.
A common implementation strategy for a reliable causal broadcast 
service~\cite{Schmuck88theuse} is to assign a vector clock to each message
broadcast, and use the causality information in the vector clock to decide at
each destination when a message can be delivered. If a message arrives at a
given destination before causally preceding messages have been delivered,
the service delays delivery of that message until those messages arrive and
are delivered. 

When messages are delivered to the application layer, in a sequence that is consistent with causality, there is some loss of information when using standard RCB middleware. Messages that the middleware knows be concurrent, are delivered in order and the application is unaware of that concurrency. Also the middleware can often determine that, for a given delivered message, that no more messages concurrent to it are to be delivered. This information, that can be used to optimize message processing by the application, is also not conveyed. Next we show how to expose this ordering information and how to provide a causal stability oracle that informs on stable messages, those with no further concurrent deliveries.

\subsection{Exposing Ordering Information}
Reliable Causal Broadcast (RCB) uses internally a logical timestamp $t$
(e.g., a vector clock) to order messages. We propose to simply expose
this information to the upper layers. Technically, the 
$\cdeliver(m)$ middleware API (used in Algo.~\ref{algo:op-rcb}) will now 
look like this: $\tcdeliver(t, m)$; meaning that, the calling process
will receive the message $m$ as well as its associated timestamp $t$.
Having a timestamp provided by the middleware allows the application logic to access causality information and avoids duplicating it in the datatype logic, avoiding the complexity in Figure \ref{fig:op-set} and simplifying the message payload.

\subsection{Causal Stability}
In concurrent data types, the order information provided by timestamps
may only be needed for an operation as long as its concurrent operations are being 
delivered or expected. However, in our experience, this information is useless
once no concurrent operations are expected for a given operation, which we then call
a ``causally stable'' operation
and thus, it makes sense to get rid of this extra meta-data.
Consequently, we propose another extension to the RCB middleware in order to 
provide \emph{causal stability} information to the upper layers. We first
define causal stability.

\begin{definition}[Causal Stability]
A timestamp $\tau$, and a corresponding message, is \emph{causally stable} at node $i$
when all messages subsequently delivered at $i$ will have timestamp $t > \tau$.
\end{definition}

This implies that no message with a timestamp $t$ concurrent with $\tau$ can
be delivered at $i$ when $\tau$ is causally stable at $i$. This notion differs
from the classical message stability, in~\cite{Birman:1991:LCA:128738.128742},
in which a message is considered stable if it has been received by all nodes.
Our definition is stronger, as we also require that no further concurrent
messages may be delivered.
Our definition is about delivered (as opposed to received) messages, and is
per-node (some $m$ may be causally stable at some node $i$ but not at some
node $j$), not global. It is similar to the definition in
\cite{rep:syn:sh138}, but we consider important to use a different name, to
avoid confusion with classic message stability.

The TCSB middleware can offer this \emph{causal stability} information through 
extending its API with $\tcstable_i(\tau)$ which informs the upper layers that
message with timestamp $\tau$ is now known to be causally stable.
This can be done by each node having a causal stability oracle which
conservatively detects causal stability according to local knowledge.
Each node $i$ can check if timestamp $\tau$ is causally stable at
$i$ by verifying if a message with timestamp $t > \tau$ has already been
delivered at $i$ from every other node $j$ in the set of nodes $I$. More
formally, if $\deliv_i()$ returns the set of messages that have been delivered
at node $i$ and $\src(t)$ denotes the node from where the message
corresponding to $t$ has been issued:
\[
  \tcstable_i(\tau)  \kw{if} \forall j \in I \setminus \{i\} \cdot
   \exists t \in \deliv_i() \cdot \src(t) = j \wedge \tau < t.
\]

One possible implementation of a TCSB causal stability oracle uses a strategy
similar to Roh's RGA tombstone deletion algorithm \cite{syn:app:1649}.
Each node $i$ keeps a local map $L_i$ (from $I$ to $T$) with the last vector
clock timestamp, delivered locally, from each other node.
We define an auxiliary function $\low$ that gives the greatest lower bound on messages
issued at $j$ delivered at each other node:

\[
  \low(L,j) \doteq \min(\{ L(k)(j) | k \in \dom(m) \}).
\]

For instance, if $\low(L_i,j)=4$, we know at node $i$ that all other nodes
have delivered at least message number 4 from node $j$.
Using this function we can now define a sufficient condition for the causal
stability oracle.  A timestamp $\tau$ is causally stable at node $i$ if
\[
  \tau(\src(\tau)) \leq \low(L_i,\src(\tau)).
\]

This ensures at $i$ that message $\tau$ was already delivered at all nodes and
that each node issued a message after delivering $\tau$, that is already
delivered at $i$. Since any messages concurrent to $\tau$ from any node $k$
must have been issued prior to $\tau$ being delivery at $k$ then, due to
causal delivery, they must have also been delivered at node $i$.

\section{Pure CRDTs Based on Partially Ordered Logs (\polog)}
\label{sec:polog}
We introduce a new framework for designing pure op-based CRDTs of
non-commutative data types. Driven by the challenges discussed in Section~\ref{sec:challenges},
the framework supports concurrent data types through the use of a Partial Ordered Log (\polog 
for short) which retains all invoked operations together with their timestamps. Being ``pure'',
i.e., the $\prepare$ in framework shall only have the name of the operation and potential arguments,
the timestamp that is stored in the \polog is directly fed by the underlying TCSB middleware
introduced above. Figure~\ref{fig:polog-reference} depicts a reference design of a pure
CRDT, whereas Algorithm~\ref{algo:ercb} describes its interaction with the TCSB middleware.
We describes the framework in this section showing that it is easy to understand
and generic. The framework is however not practical without the optimizations we introduce
in the subsequent sections --- due to the ever-growing \polog.

The challenges leveraged in Section~\ref{sec:challenges} can be summarized by 
two points. The first is that there is a real need to support data types that
are concurrent in practice. Consequently, we choose to retain all operations in a partially 
ordered log (\polog) that can be implemented as a map from timestamps to 
operations: $\Sigma:=T\map O$ as shown in Figure~\ref{fig:polog-reference}. 
This brings two main benefits: (1) a \polog can retain 
concurrent operations, and (2) storing operations as black-boxes makes the 
state type  generic across all data types.
The timestamps offered by the TCSB middleware through $\tcdeliver$ (as in 
Algorithm~\ref{algo:ercb}) are crucial to keep information 
about concurrent operations, not trying to impose a local total-order over 
them, contrary to a classic sequential log.

The second challenge, i.e., how to design the $\effect$, is interestingly made
easy using the \polog.
In fact, we can now obtain a universal data-type-independent definition of $\effect$:
\[
\effect(o, t, s) = s \union \{(t, o)\},
\]
which associates a timestamp to the corresponding operation and joins it to the \polog
via the set union $\union$, which is commutative, therefore making the $\effect$ commutative
as needed.

In this novel framework, it is easy to notice that all the components, but one,
of the data type design are the same regardless of the data type. This significantly 
reduces the burden of implementation on developers as well as the chances of
introducing new bugs.
Only the query functions, defined in $\eval$, will need to be data-type-specific 
according to the desired semantics; however, their definition over the 
\polog will typically be a direct transposition of their specification, which 
is fairly easy to define as we show next in the examples.

\begin{figure}[t]
\begin{eqnarray*}
\Sigma = T \map O &&
\sigma_i^0 = \{\}  \\
\prepare_i(o, s) & = & o \\
\effect_i(o, t, s) & = &  s \cup \{(t,o)\} \\
\eval_i(q, s ) & = & \textrm{[datatype-specific query function]} \\
\end{eqnarray*}
\caption{\polog based reference implementation for pure op-based CRDTs}
\label{fig:polog-reference}
\end{figure}

\begin{algorithm}[t]
\DontPrintSemicolon
\SetKwBlock{state}{state:}{}
\SetKwBlock{period}{periodically:}{}
\SetKwBlock{on}{on}{}
\state{
  $\sigma_i \in \Sigma$ \;
}
\on({$\operation_i(o)$:}){
	$m:=\prepare_i(o, \sigma_i)$ \;\\
	$\tcbcast_i(m)$ \;
}
\on({$\tcdeliver_i(m, t)$:}){
  $\sigma_i := \effect_i(m, t, \sigma_i)$ \;
}
\BlankLine
\on({$\tcstable_i(t)$:}){
  $\sigma_i := \stable_i(t, \sigma_i)$ \;
}
\BlankLine
\caption{Distributed algorithm for node $i$ showing 
	the interaction between the Tagged
Causal Stable Broadcast (TCSB) middleware and a Pure CRDT.}
\label{algo:ercb}
\end{algorithm}

\subsection{Example CRDTs}
We exemplify the use and simplicity of the above framework through two
examples: an Add-Wins Set (\awset) and a Multi-Value
Register (\mvreg). In a nutshell, the semantics of \awset give priority for
$\add$ operation over a $\remove$ if they are concurrent; whereas, an \mvreg
retains all concurrently written values and leaves the decision to the
application to choose the desired value. These semantics are explained in
details in Section~\ref{sec:catalog}.

Figures~\ref{fig:polog-orset} and~\ref{fig:polog-mvreg} show two pure \polog
designs for \awset and \mvreg, respectively, using the framework presented
in Fig.~\ref{fig:polog-reference}. It is interesting to notice that the pure
CRDT design of both data types is exactly the same considering the state $\Sigma$,
$\prepare$, and also $\effect$. The only data-type-specific function is the
$\eval$ query function which defines the behavior of the data types in both
CRDTs. 
In \awset (Fig.~\ref{fig:polog-orset}), the values reported to be in
the set are those corresponding to the $\add$ operations in the \polog 
having no $\remove$ in their causal future. 
Whereas, in \mvreg (Fig.~\ref{fig:polog-mvreg}), a read
reports the set of all distinct concurrently written values, via $\af{wr}$, that have not
been overwritten.

\begin{figure}[t]
\begin{eqnarray*}
\Sigma = T \map O &&
\sigma_i^0 = \{\}  \\
\prepare_i(o, s) & = & o
\qquad\qquad\qquad (o \textrm{ is } [\add, v] \textrm{ or
} [\rmv, v])\\
\effect_i(o, t, s) & = &  s \cup \{(t,o)\} \\
\eval_i(\elems, s ) & = & \{ v | (t, [\add,v])
\in s \land \not\exists (t',[\rmv,v]) \in s \cdot t < t'\}\\
\end{eqnarray*}
\caption{\polog based Add-Wins Set (\awset).}
\label{fig:polog-orset}
\end{figure}

\begin{figure}[t]
\begin{eqnarray*}
\Sigma = T \map O &&
\sigma_i^0 =  \{\}  \\
\prepare_i(o, s) & = & o
\qquad\qquad\qquad (o \textrm{ is } [\af{wr}, v])\\
\effect_i(o, t, s) & = &  s \cup \{(t,o)\} \\
\eval_i(\rd, s ) & = & \{ v \mid (t, [\af{wr},v])
\in s \land \not\exists (t',[\af{wr},v']) \in s \cdot t < t' \}\\
\end{eqnarray*}
\caption{\polog based Multi-Value register (\mvreg)}
\label{fig:polog-mvreg}
\end{figure}

Although these designs are not realistic to be used in practice, since the
state size in each replica grows linearly with the number of operations, they
represent a starting point from which actual efficient designs can be derived by
semantic \polog compaction, as we show in the rest of the paper. In addition,
these designs are theoretically relevant as they provide a clear description
of the concurrent semantics of the replicated data type, and therefore can be used
as a base-line to compare with other designs. This is made possible
through capturing the partial ordered set of all operations delivered to each
replica\footnote{A similar approach to express data type semantics is found in
\cite{DBLP:conf/popl/BurckhardtGYZ14} when relating to the \emph{visibility}
relation.}.

\section{Semantic \polog Compaction}
\label{sec:compaction}

In this section, we show how \polog based CRDTs can be made practical by
performing \polog compaction, and thus extend the introduced framework.  The
purpose of compaction is to reduce the space and computation overheads through
pruning the \polog in a lossless manner. We do this through two main
mechanisms: \emph{causal redundancy} which once an operation is delivered, one 
may prune that very operation or/and an already existing \polog operation,
whichever is deemed redundant according to the data type semantics; and
\emph{causal stabilization} which once some operation in the \polog becomes
causally stable, it discards the operation timestamp and possibly removes it or
even some other operations.

\subsection{Causal Redundancy}

The first mechanism prunes the \polog once an operation is
causally delivered in the $\effect$. The aim is to keep the smallest number of \polog 
operations such that all queries return the same result as if the full \polog was
present. In particular, this method discards operations from the \polog if 
they can be removed without impacting the output of query operations.
Since this pruning is based on the semantics of a data type, it is obviously 
data-type-dependent; however, we opt to call it causal redundancy
since pruning is mainly driven by causality.

In Fig.~\ref{fig:compact-framework}, we extend the generic framework
which now includes the \polog, $\prepare$, and a more sophisticated generic
$\effect$. The $\effect$ can now prune the \polog through three data-type-specific
relations that together define causal redundancy.
This has the advantage of keeping the framework generic, with the same
$\effect$ for all data types, while delegating data-type-specific behavior to
mere relations (and not arbitrary procedures).

\begin{figure}[t]
\begin{eqnarray*}
\Sigma = T \map O &&
\sigma_i^0 = \{\}  \\
\prepare_i(o, s) & = & o \\
\effect_i(o, t, s) & = &
  \{(t,o) | (t,o) \not\R s\} \cup
  \{x \in s | x \not\R_0 (t,o) \\ && \land (t,o) \R s \lor
              x \not\R_1 (t,o) \land (t,o) \not\R s\} \\
\R, \R_0, \R_1 & = & \textrm{[datatype-specific redundancy relations]} \\
\eval_i(q, s ) & = & \textrm{[datatype-specific query function]} \\
\end{eqnarray*}
\caption{Generic framework with \polog compaction by causal redundancy}
\label{fig:compact-framework}
\end{figure}

The $\R$ relation, in Figure~\ref{fig:compact-framework}, defines whether the 
delivered operation is itself redundant
and does not need to be added itself to the \polog. In most cases, as 
we will see later, this can be decided by looking only at the delivered operation 
itself, regardless of the current \polog and, thus, a unary $\R$ would be enough;
however, in other cases, this depends on the \polog; therefore, $\R$ is a binary 
relation between the operation and the \polog.

The other two relations, $\R_0$ and $\R_1$, define which operations in the
current \polog become redundant given the delivery of the new operation. For
some data types this decision depends on whether the delivered operation is
added to the \polog or whether it is itself redundant: we may be able to
discard some operation if the new one is kept, but not if it is itself
discarded. Having these two relations makes the framework expressive enough
for such data-types (which are fortunately not numerous): 
$\R_0$ is used when the new arrival is discarded 
being redundant, and $\R_1$ if it is added to the \polog. 
For most data types, these relations will be equal, i.e., $\R_0 = \R_1$, 
and we define a single $\R_\_$ relation to denote both $\R_0$ and $\R_1$.
As we will see, the \polog compaction resulting from these relations will
provide invariants over the \polog which allow simplified query functions that
give the same result over the compact \polog as the original query functions
over the full \polog.

Having explained the roles of $\R$, $\R_0$, and $\R_1$, the behavior of $\effect$
in Figure~\ref{fig:compact-framework} becomes as follows: a newly delivered operation
$(t,o)$ is added to the \polog if it is not redundant by the \polog operations,
as the first clause says: $\{(t,o) | (t,o) \not\R s\}$. The remaining clause says
that an existing operation $x$ in the \polog is removed if it is made redundant by $(t,o)$. 
Since, the behavior is sometimes different depending on whether $(t,o)$ has been discarded 
or not, the two clauses, separated by $\lor$, distinguishes these two cases.

\subsubsection{Example CRDTs}

We apply this \polog compaction to the previous examples presented in
Figures~\ref{fig:polog-orset} and~\ref{fig:polog-mvreg}, extended with a $\clear$ operation that removes all values; the corresponding
compact designs are presented in Figures~\ref{fig:or-set-compacted} 
and~\ref{fig:mvreg-compacted}, respectively. In the latter two, we
only mention the data-type-specific functions since $\Sigma$, $\prepare$, 
and $\effect$ are generic. In these examples,
the left side of the relations $\R$ and $\R_\_$ are made redundant by the
right side. On the other hand, the notation $o[i]$ refers to the $i^{th}$
element in a list that comprises the operation name and subsequent arguments (analogous to
the $argv[]$ used in several programming languages); in particular, $o[0]$ 
refers to the operation name.

In the \awset design in Figure~\ref{fig:or-set-compacted}, the first clause says
that $\clear$ and $\remove$ operations are always made redundant by
(and thus not stored in) the \polog. The second clause specifies that an existing
operation $o'$ in the \polog is redundant (and thus pruned) if the newly arrived 
operation $o$ is strictly in its causal future, provided that $o$ is either $\clear$
or otherwise (e.g., $\add$) tries to modify the same value (i.e., set item) in the set (in 
which case it is clearly redundant). These relations will actually result in a \polog
that is deprived of any $\rmv$ or $\clear$ operations; consequently, the $\af{rd}$
operation will simply return all the values in the \polog.

The second example is the \mvreg, presented in Figure~\ref{fig:mvreg-compacted}.
The first clause specifies that a $\clear$ operation is always redundant and must not be
added to \polog. The second clause says that an existing operation $o'$ is made redundant
by $o$ if the latter is strictly in its causal future; otherwise all concurrent $\af{wr}$
operations are kept in the \polog. The $\af{rd}$ operation is thus very simple: it returns
all the present values in the \polog.

\begin{figure}[t]
\begin{eqnarray*}
	(t,o) \R s &\iff& o[0] = \clear \lor \rmv \\
	(t',o') \R_\_ (t,o) &\iff& t' < t \land (o[0] = \clear \lor o[1] = o'[1]) \\
  \eval_i(\elems, s) & = & \{ v | (\_, [\add,v]) \in s \}\\
\end{eqnarray*}
\caption{Add-Wins set with \polog compaction. $o[0]$ denotes the operation name
and $o[1]$ denotes its first argument.}
\label{fig:or-set-compacted}
\end{figure}

\begin{figure}[t]
\begin{eqnarray*}
	(t,o) \R s &\iff& o[0] = \clear \\
  (t', o') \R_\_ (t, o) &\iff& t' < t \\
  \eval_i(\af{rd}, s) & = & \{ v | (\_, [\af{wr}, v]) \in s \}\\
\end{eqnarray*}
\caption{Multi-Value register with \polog compaction}
\label{fig:mvreg-compacted}
\end{figure}

\subsection{Causal Stabilization}

\emph{Causal stabilization} is the second \polog compaction mechanism; it
exploits \emph{causal stability} information to discard timestamp
meta-data for the elements that become \emph{causally stable}, and may allow
some elements to be entirely removed from the \polog.

\subsubsection{Discarding stable timestamps}
This mechanism discards the causally stable timestamps in the \polog.
The motivation is that the $\effect$ function makes use of the relations $\R$, 
$\R_0$, and $\R_1$ to compare $(t',o')$ elements in the \polog with newly 
delivered operations, but never compares \polog elements among themselves.
Our observation is that most of the meta-data in the \polog is not
needed when incoming operations are no longer concurrent.
Considering the definition of causal stability in Section~\ref{sec:ercb}:
if some pair $(t',o')$ is in the \polog, with $t'$ a causally stable timestamp,
all future deliveries $(t, o)$ used in $\effect$ will be in its causal 
future, i.e., $t' < t$. This means that what is important is to know if this 
comparison $t' < t$ is true or false regardless of the exact timestamp values.
Consequently, we can replace an existing timestamp $t'$ that became stable with
another one that is always less than $<$ any other timestamp $t$ to be 
delivered. (Notice that $t$ cannot be replaced now until it also becomes stable,
otherwise the concurrent semantics will break.) For instance, a
stable timestamp $t'$ can be replaced by $\bot$, i.e., the least element (a.k.a., 
bottom) of the timestamp domain, effectively discarding the timestamp; e.g.,
$0$ is the $\bot$ of a scalar timestamp in $\nat$.

However, to perform such a timestamp replacement, two necessary conditions 
have to be met in the data-type-specific functions: queries should not be
affected and the causal redundancy relations $\R$, $\R_0$, and $\R_1$ must
not compare the existing \polog timestamps together. It was a fortunate happenstance that
this was the case for all data types that we have considered in this paper, and
thus all can be causally compacted using stabilization.
The explanation of this is that, in all these data types, after causal 
redundancy is applied, the resulting simplified query operations no longer 
need to compare timestamps, and therefore, the queries will
give the same result as before even if stable timestamps are replaced by $\bot$.
On the other hand, $\R$, $\R_0$ and $\R_1$ never compare existing timestamps in the 
\polog with each other, but only compare them with the newly arrived operations.

This optimization can have substantial gains in many data types as it 
greatly diminishes the size of a replica state.
In practice, instead of having timestamps that are maps or
vectors with linear size on the number of replicas, we can have a special
marker denoting $\bot$ (e.g., a null pointer). Consequently, we designate two 
interesting data type classes. 
The first class is those CRDTs where values may take considerably less space 
than timestamps, like sets of integers, and thus, stripping the timestamp from an 
element can reduce the storage overhead several orders of magnitude.
The second class is those CRDTs whose state size can grow significantly;
in such CRDTs, the percentage of elements in the \polog that are not yet causally 
stable will be quite small, most of operations being already stabilized. 
Therefore, this optimization reduces the storage overhead of the element itself,
and may impact most \polog elements, as they become causally stable.

\emph{Further optimizations in practice:}
In actual implementations, the \polog can be split in two components: 
one that simply stores the set of stable operations and the other stores the 
timestamped operations.
For instance, instead of being a map $T \map O$, the state is split into
\[\Sigma=\pow{O} \times (T \map O)\]
detaching into a plain set of all stable operations. Furthermore, for some
data types (such as \awset) where only one kind of operations, i.e., $\add$, 
is ever in the \polog, the stable set of operations can become a plain
set of elements (without timestamps), 
e.g., the traditional sequential set data type, with possibly
specialized implementations according to the domain (i.e., a bitmap for dense
sets of integers). To keep the presentation clear and consistent, we leave 
such optimizations to actual implementations and we rather only consider 
using a unified \polog, containing both stable $(\bot, o')$ and unstable operations $(t', o')$. 

\subsubsection{Discarding stable operations}
This mechanism allows to discard stable operations, not only timestamps, if
they have no impact on the semantics of the data type.
The idea is that, for some data types like \rwset (explained next), 
we noticed that some operations
could not be considered redundant once delivered, but they become useless once
other future operations become stable; and therefore,
it may be possible to discard a set of operations at once. \\

To support both causal stabilization mechanisms, we extend the 
framework, presented in Figure~\ref{fig:full-compaction}, to include a 
data-type-specific $\stabilize$ function. This function takes a stable timestamp 
$t$ (fed by the TCSB middleware) and the full \polog $s$ as input, 
and returns a new \polog (i.e., a map), possibly discarding a set of
operations at once. The $\stable$ handler, on the other hand, invokes $\stabilize$
and then strips the timestamp (if the operation has not been discarded by
$\stabilize$), by replacing a $(t',o')$ pair that is present in the returned \polog by
$(\bot, o')$. In the following section, we show how the \awset and \mvreg, presented in 
Figures~\ref{fig:or-set-compacted} and~\ref{fig:mvreg-compacted},
as well as other several data types, can be optimized with causal stabilization.

\begin{figure}[t]
\begin{eqnarray*}
\Sigma = T \map O &&
\sigma_i^0 = \{\}  \\
\prepare_i(o, s) & = & o \\
\effect_i(o, t, s) & = &
  \{(t,o) | (t,o) \not\R s\} \cup
  \{x \in s | x \not\R_0 (t,o) \\ && \land (t,o) \R s \lor
              x \not\R_1 (t,o) \land (t,o) \not\R s\} \\
  \stable_i(t, s) &=& \stabilize_i(t, s)[ (\bot,o) / (t,o)] \\
\R, \R_0, \R_1 &=& \textrm{[datatype-specific redundancy relations]} \\
  \stabilize_i(t,s) &=& \textrm{[datatype-specific stabilization function]} \\
\eval_i(q, s ) & = & \textrm{[datatype-specific query function]} \\
\end{eqnarray*}
\caption{Pure CRDTs framework including \polog compaction: causal redundancy and stabilization.}
\label{fig:full-compaction}
\end{figure}

\section{Portfolio of Pure Op-based CRDTs}
\label{sec:catalog}
In this section, we provide a catalog of CRDT designs following the pure
op-based framework introduced above. Our purpose it to make this framework and
CRDTs easy to understand and implement by designer and developers.
We try to cover examples of the most 
used data types in practice, e.g., in Dynamo~\cite{app:rep:optim:1606}, 
Riak~\cite{riak-crdt:Cribbs:2012}, 
Cassandra~\cite{cassandra:2016}, etc. In particular, we address two 
categories of data types: (1) Commutative CRDTs: Grow-only Counter (\gctr),
Positive-Negative Counter (\pnctr), Grow-only Set (\gset), and Two-Phase Set 
(\twopset); and Non-commutative data types: Enable-Wins Flag (\ewfl), Disable-Wins 
Flag (\dwfl), Multi-Value Register (\mvreg), Add-Wins Set (\awset), and 
Remove-Wins Set (\rwset). To limit redundancy, but still provide reference specifications, the \gctr and both flags are collected in appendix.

In this catalog, our methodology is to define the data type
and show how it can be used in practice. Then we describe its semantics in a 
generic way before presenting its pure op-based design. This analogy helps
the designer/developer choose the desired semantics for his business logic as well as 
understand the design of the CRDT in the Pure framework.
Finally, we discuss additional optimizations that are data type
specific when useful. (Recall that the notation used in this section are
described in Section~\ref{sec:model}). 


\subsection{Commutative Data Types}
According to Section~\ref{sec:pure}, a CRDT is said to be commutative if all 
its possible operations are commutative over the state.
The design of commutative CRDTs is datatype-specific, both in 
the pure op-based as well as in classical op-based approaches. Indeed, it 
is possible to design them using the \polog framework to be generic across all data types,
as we do for other non-commutative data types; however, the tradeoff is to induce an 
extra unneeded overhead. We opt not to use the \polog for such
data types and thus provide more practical designs. All the commutative designs
assume the existence of a Reliable Causal Broadcast (RCB) middleware. The interaction
between the presented CRDTs and the RCB are discussed in Algorithm~\ref{algo:op-rcb},
Section~\ref{sec:back}.

%

\subsubsection{Positive-Negative Counter (\pnctr)}
A \pnctr is a counter that supports increment and decrement operations. 
Therefore, it is very useful to represent quantities that can be added or removed, like
the number of available tickets in a system.
Its state is usually implemented as $n\in \mathbb{Z}$ with two operations 
$\inc$ ($+$ operator) and $\dec$ ($-$ operator); thus, its value can go negative. 
Since $+$ and $-$ are commutative in $\mathbb{Z}$, \pnctr is a commutative data type.

Figure~\ref{fig:pnctr} conveys the pure CRDT design of \pnctr. 
The state is now defined as integer $z\in \mathbb{Z}$ that is initialized to $0$.
$\prepare$ can be either $\inc$ or $\dec$; however, in both cases, no meta-data is
needed (and thus it is pure). The $\effect$ function increments or decrements $z$ 
depending on $\prepare$'s output. Finally, $\eval$ invokes the read operation $\af{value}$ and
simply returns $z$.

\begin{figure}[t]
\centering
\begin{eqnarray*}
\Sigma = \mathbb{Z}&&
\sigma_i^0 = 0 \\
\prepare_i(\inc, z) & = & \inc \\
\prepare_i(\dec, z) & = & \dec \\
\effect_i(\inc, z ) & = & z+1 \\
\effect_i(\dec, z ) & = & z-1 \\
\eval_i(\af{value},z ) & = & z\\
\end{eqnarray*}
\caption{Pure Positive-Negative Counter CRDT (\pnctr).}
\label{fig:pnctr}
\end{figure}

\subsubsection{Grow-only Set (\gset)}
A \gset is a set where only adding items is supported. It is usually used when removals
are not needed, as in storing the IPs of clients visiting a web-site.
The set comprises a collection of items, of some type, that are added via 
$\add$ operation (a union $\union$). Since $\add$ is commutative over the set, 
\gset is a commutative datatype. 

Figure~\ref{fig:gset} shows the pure CRDT design of \gset.
The state is represented by $s\in \pow{V}$ (a power set of $V$) and is originally empty. 
The function $\prepare$ can only be the $\add$ operation with its parameter $v$, but
without extra meta-data (which makes it pure). The function $\effect$ adds $v$ to 
the set using the set union operator: $\union$. Finally, the $\eval$ function invokes
the read operations $\elems$ and $\size$ to return all the elements of the set or 
its cardinality, respectively.

\subsubsection{Two-Phase Set (\twopset)}
The \twopset is a set where items can be added and removed, but never 
added after being removed. It can be seen as a composition of two 
\gset, one for additions and another for removals. 
In practice, the \twopset can be used to represent collections having
items with unique ids. The allowed mutating operations in \twopset are 
$\add$ and $\rmv$. They are commutative since the presence of an item in
the removals set  prevents re-adding that element later;
Both $[\add,a]$ followed by $[\rmv,a]$, and $[\rmv,a]$ followed by $[\add,a]$
will have the same effect, i.e., $a$ not being present.
Therefore --- contrary to what it looks like at a glance --- a \twopset is a commutative CRDT.

Figure~\ref{fig:2pset} depicts the pure CRDT design of the \twopset. The state
is designed as two power sets $s,p\in \pow{V}$, initially empty, where $s$ retains 
the added values and $p$ retains the removed ones. A $\prepare$ can either be 
$\add$ or $\rmv$ together with a parameter $v$ and without additional meta-data
(i.e., making the CRDT pure). When item $v$ is added, the $\effect$
unions $v$ to set $s$ only of it is not in the set of removals $p$ 
(i.e., $s \union (\{v\} \setminus p)$), which remains intact in this case. 
If $v$ is to be removed via $\rmv$, it is subtracted from $s$ and 
unioned to $p$. Notice that, set $s$ can grow and shrink in 
size depending on the issued operations whereas $p$ can only grow in size as long 
as $\rmv$ operations are issued. This imposes a space overhead, but ensures that removed
items are never added again. Finally, the $\eval$ function invokes
the read operations $\elems$ and $\size$ to return all the elements in $s$ or 
its cardinality, respectively.

\begin{figure}[t]
\begin{subfigure}[b]{.50\textwidth}
\centering
\begin{eqnarray*}
\Sigma = \pow{V} &&
\sigma_i^0 = \{\} \\
\prepare_i([\add, v], s) & = & [\add, v] \\
\effect_i([\add, v], s) & = & s \union \{v\} \\
\eval_i(\elems, s) & = & s\\
\eval_i(\size, s) & = & \setsize{s}\\
\end{eqnarray*}
\caption{GSet}
\label{fig:gset}
\end{subfigure}
\begin{subfigure}[b]{.50\textwidth}
\centering
\begin{eqnarray*}
\Sigma = \pow{V}\times \pow{V} &&
  \sigma_i^0 = (\{\},\{\}) \\
\prepare_i([\add, v], \sigma) & = & [\add, v] \\
\prepare_i([\rmv, v], \sigma) & = & [\rmv, v] \\
  \effect_i([\add, v], (s,p)) & = & (s \union (\{v\} \setminus p),p) \\
  \effect_i([\rmv, v], (s,p)) & = & (s\setminus \{v\},p \union \{v\}) \\
  \eval_i(\elems, (s, p)) & = & s \\
  \eval_i(\size, (s,p)) & = & \setsize{s}\\
\end{eqnarray*}
\caption{2PSet}
\label{fig:2pset}
\end{subfigure}
\caption{Pure Grow-only Set and Two-Phase Set CRDTs.}
\label{fig:pure-set}
\end{figure}

\subsection{Non-Commutative Data Types}
These CRDTs accept operations that are not commutative. 
The challenge is
then to design the $\effect$ function to be commutative.
The design of such data types follows the framework presented in 
Figure~\ref{fig:full-compaction} and supports using the \polog as well as the
compaction mechanisms explained in the preceding sections. Consequently, the
presented pure CRDTs have a common design for the state $\Sigma$, 
$\prepare$, $\effect$, and $\stable$ functions. 
Although the other functions like $\R$, $\stabilize$, and $\eval$ are datatype-specific,
they are often fairly simple as shown next.
As explained earlier, non-commutative pure CRDTs assume the existence of an  
Tagged Causal Stable Broadcast middleware (TCSB), as presented in Section~\ref{sec:ercb}.
The interplay between the TCSB and the pure CRDTs is described in Algorithm~\ref{algo:ercb}.

Since the semantics of non-commutative CRDTs is more complex than
commutative ones, being concurrent, we formally define their concurrent semantics assuming
that all operations are retained in a \polog, where the output of the query operations
defines its semantics. This makes the definition very solid and stands as a base-line to verify 
future amendments across; i.e., the output of defined queries for a datatype must not change in 
any new design or optimization (which is true in all the designs we present).

\subsubsection{Multi-Value Register (\mvreg)}
Contrary to a classical register that holds a single value at a time, 
a \mvreg can hold multiple values if their corresponding $\af{write}$ 
operations are concurrent;
thus letting the application choose the desired values returned.
One motivation for \mvreg was to mimic the design of the Amazon Shopping Cart in
Dynamo DB~\cite{app:rep:optim:1606}. Since the order of two $\af{write}$ operations
on the \mvreg can lead to different results if one of them is in the causal 
future of the other, the datatype is non commutative. 

The concurrent
semantics of \mvreg is depicted in Figure~\ref{fig:mvreg-semantics} and says that a read $\af{rd}$
should return all write $\af{wr}$ operations that are not causally succeeded by any other 
operation (i.e., neither $\af{wr}$ nor $\clear$). Consequently,
the first clause in the Pure CRDT design, in Figure~\ref{fig:mvreg-compact}, 
says that $\clear$ operations must not be added to the \polog. The second 
clause then means that $\af{wr}$ operations are removed from the \polog if there
is any $\af{wr}$ or $\clear$ operation in the causal future; otherwise, 
concurrent $\af{wr}$ operations will be kept. The $\af{rd}$ operation, 
therefore, simply returns all available $\af{wr}$ operations in the \polog.
Finally, $\stabilize$ has no effect on the \polog, and thus the calling function
$\stable$ will simply replace causally stable timestamps with $\bot$.

\begin{figure}[t]
	\begin{eqnarray*}
  \eval_i(\af{rd}, s) & = &
  \{ v | (t, [\af{wr}, v]) \in s \land \forall (t', \_) \in s \cdot t \not < t' \}
\end{eqnarray*}
\caption{\mvreg concurrent semantics}
\label{fig:mvreg-semantics}
\end{figure}
\begin{figure}[t]
\begin{eqnarray*}
	(t,o) \R s &\iff& o[0] = \clear \\
  (t', o') \R_\_ (t, o) &\iff& t' < t \\
  \stabilize_i(t,s) &=& s \\
  \eval_i(\af{rd}, s) & = & \{ v | (\_, [\af{wr}, v]) \in s \}\\
\end{eqnarray*}
\caption{Compact \mvreg CRDT}
\label{fig:mvreg-compact}
\end{figure}

\subsubsection{Sets (\awset and \rwset)}
A set is a very well known data structure that retains unordered items, being 
added and removed through $\add$ and $\rmv$ operations, respectively. 
Obviously, these operations are not commutative and thus reordering them
will yield different effects, making the datatype non commutative.
Special consideration must be given when
these operations are concurrent. Typically, the designers explicitly define
extra rules to make the behavior consistent across different replicas, thus
leading to two variants: Add-Wins Set (\awset) and Remove-Wins Set (\rwset).

In \awset, an $\add$ operation dominates a concurrent $\rmv$
that is mutating the same item in the set. 
This is explained formally in the concurrent semantics in
Figure~\ref{fig:awset-semantics}: a $\elems$ operation will return all the values
associated with $\add$ such that no corresponding $\rmv$ operation
or $\clear$ operation are found in the causal future of the $\add$. 
Consequently, the first clause in the Pure CRDT design in 
Figure~\ref{fig:awset-compact} specifies that newly delivered 
$\rmv$ and $\clear$ operations are never added to the \polog; 
which leaves the \polog with only $\add$ operations.
The second clause however says that an existing $\add$ in the \polog is
considered redundant if there is another $\add$ or $\rmv$ operations on 
the same item (specified by $o[1]$ in Figure~\ref{fig:awset-compact}) 
or a $\clear$ operation such that they are in the 
causal future of the candidate $\add$. Therefore, the $\elems$ operation
will simply return the items associated with an $\add$ in the \polog, since
neither duplicate $\add$s nor $\rmv$ and $\clear$ operations exist in the \polog.
For the same reasons as previous data types, $\stabilize$ has no effect and
the calling $\stable$ function will only replace the causally stable timestamps 
with $\bot$.

\begin{figure}[t]
\begin{eqnarray*}
  \eval_i(\elems, s) & = &
  \{ v | (t, [\add, v]) \in s \\&&\land 
    \forall (t', [\rmv, v] \lor \clear) \in s \cdot t \not < t' 
\}
\end{eqnarray*}
\caption{\awset concurrent semantics}
\label{fig:awset-semantics}
\end{figure}
\begin{figure}[t]
\begin{eqnarray*}
	(t,o) \R s &=& o[0] = (\clear \lor \rmv) \\
	(t',o') \R_\_ (t,o) &=& t' < t \land (o[0] = \clear \\
	&&\lor o[1] = o'[1]) \\
  \stabilize_i(t,s) &=& s \\
  \eval_i(\elems, s) & = & \{ v | (\_, [\add,v]) \in s \}\\
\end{eqnarray*}
\caption{Compact \awset Pure CRDT}
\label{fig:awset-compact}
\end{figure}

On the other hand, in \rwset, a $\rmv$ operation dominates a 
concurrent $\add$ that is mutating the same item in the set, thus masking its
effect. Therefore, the formal concurrent semantics in Figure~\ref{fig:rwset-semantics}
says that a $\elems$ operation on a \polog returns all values associated
with an $\add$ such that all $\rmv$ operations are in its causal past, and
$\clear$ operations are not in the causal future. Notice that, contrary to 
\awset, concurrent $\add$ operations with a $\rmv$ will be discarded (reflecting
the remove-wins semantics). The corresponding Pure CRDT design of \rwset is
presented in Figure~\ref{fig:rwset-compact}. In particular, the first clause
prevents any $\clear$ operation from being added to \polog. The second clause
specifies that a candidate operation on item $v$ is only removed if there is another
operation on $v$ or a $\clear$ in its causal future. This actually means that
$\rmv$ operations can still live in the \polog, to the contrary of \awset.
In fact, $\rmv$ cannot be removed since a concurrent $\add$ on the same
item may be delivered which breaks the remove-wins semantics. Even more,
an $\add$ operation $a$ cannot be removed despite the presence of concurrent 
$\rmv$ operation $r$ (as one might think). 
The reason is that $r$ can become stable at any time; if it happened to be followed
by a $\clear$ operation, i.e., in its causal future, all stable operations in 
the \polog will be deleted, $r$ among them. Since $a$ is not yet stable, and thus 
concurrent with $\clear$, the former will stay in the \polog. This breaks the 
remove-wins semantics since $r$ and $a$ were originally concurrent and hence 
$r$ should win. These reasons lead to a more complex
$\stabilize$ function (explained next) that is in charge of removing the extra 
operations that were not safe to be removed previously through $\R$ or $\R_\_$.

\def\sepCl{\hspace{1.5pt}}
\def\s#1{\setlength{\fboxsep}{1.2pt}\sepCl\fbox{\smash{#1}\vphantom{x}}\sepCl}
\def\n#1{\dot{\s{#1}}}
  \def\c#1{\cancel{\smash{#1}\vphantom{x}}}

\begin{figure}[t]
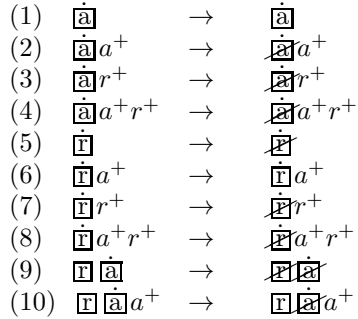

  \[
\begin{array}{l@{\quad\rightarrow\qquad}l}
 (1)\quad \n a & \n a \\
 (2)\quad \n a a^+ & \c{\n a} a^+ \\
 (3) \quad \n a r^+ & \c{\n a} r^+ \\
  (4) \quad \n a a^+ r^+ & \c{\n a} a^+ r^+ \\
  (5)\quad \n r & \c{\n r} \\
  (6)\quad \n r a^+ & \n r a^+ \\
  (7)\quad \n r r^+ & \c{\n r} r^+ \\
  (8)\quad \n r a^+ r^+ & \c{\n r} a^+ r^+ \\
  (9)\quad \s r \n a & \c{\s r} \c{\n a} \\
  (10)\ \ \s r \n a a^+ & \s r \c{\n a} a^+ \\
\end{array}
  \]
\caption{Simulation of the possible transitions at stabilization in \rwset.
The operations $\add$ and $\rmv$, i.e., $r$ and $a$ respectively, 
are assumed to mutate the same item. 
A boxed operation is stable. A dotted operation refers to the operation that has
just became stable. The sign + is used to indicate the presence of duplicate 
operations.
The canceled boxed operations on the right side are the redundant operations 
due to causal stabilization.
}
\label{fig:rwset-stabilization}
\end{figure}

We explain the $\stabilize$ function, depicted in Figure~\ref{fig:rwset-compact},
with the help of a simulation of the \polog state transitions upon stabilization in 
Figure~\ref{fig:rwset-stabilization}. The $\stabilize$ function tries to remove
redundant operations once invoked by $\stable$ when a timestamp $t$ becomes stable.
The removed operations are defined in three clauses, i.e., those removed from the 
\polog using the \emph{set minus} operator in Figure~\ref{fig:rwset-compact}.
The first clause specifies that, once $t$ becomes stable, the associated $\add$
operations are removed from the \polog as long as there is at least another 
operation in the \polog with a different timestamp $t'\neq t$ (otherwise the latter
operation will also be stable). This clause corresponds to the first four cases in
Figure~\ref{fig:rwset-stabilization}: once an $\add$ operation $a$ becomes 
stable, the existence
of other $\add$s will make it redundant, as they add the same item to the set; $a$ is
also removed if there exist any other $\rmv$ operation $r$ on the same item since,
according to the remove-wins semantics,
$r$ wins the concurrent $a$ which has just became stable, and thus, it is not expected 
to have another concurrent $\add$ delivered (by definition of causal stability).
The second clause in $\stabilize$ (Figure~\ref{fig:rwset-compact}) deletes 
the stable $\rmv$ operations as long as the potential non stable operations on 
the same item are not uniquely $\add$ operations. This is explained in the simulation
in Figure~\ref{fig:rwset-stabilization} in cases (5) through (8): cases (7) and (8)
are trivial since a $\rmv$ is simply redundant by duplicate $\rmv$s; case (5)
removes a stable $\rmv$ if no other operation exist. The rational is that the stable
$\rmv$ being stable,
no concurrent operations are expected, meaning that any delivered operations will be 
in its causal future, and thus, making stable $\rmv$ useless. Case (6) does not
lead to any removals. Indeed, if the stable $\rmv$ operation $r$ is removed,
any existing $\add$ operation on the same item $a$ will win, which contradicts with
the remove-wins semantics since $r$ and $a$ are originally concurrent. This practically
means the stable $\rmv$ will be kept in the \polog for a later stage, thus forking
the cases (9) and (10). When an $\add$ operation becomes stable, i.e., all $\add$ and
$\rmv$ operations on the same item are not stable, they can both be removed from the
\polog since $\rmv$ wins, and it is not expected to have any concurrent operation
delivered any more. To the contrary, if there is at least another $\add$ operation,
the stable $\add$ will be removed while the stable $\rmv$ will be kept for similar
reasons as in point (6) explained above.

\begin{figure}[t]
\begin{eqnarray*}
  \eval_i(\elems, s) & = &
  \{ v | (t, [\add, v]) \in s \\&&\land 
  \forall (t', [\rmv, v]) \in s \cdot t'< t \\&& \land
  \forall (t'', \clear) \in s \cdot t \not < t'' 
  \}
\end{eqnarray*}
\caption{\rwset concurrent semantics}
\label{fig:rwset-semantics}
\end{figure}
\begin{figure}[t]
\begin{eqnarray*}
	(t,o) \R s &=& o[0] = \clear \\
	(t',o') \R_\_ (t,o) &=& t' < t \land (o[0] = \clear \lor o[1] = o'[1])\\
  \stabilize_i(t,s)
  &=& s \setminus \\&& \{(t,[\add,e]) \in s |\exists (t',[\_,e])\in s \cdot t\neq t'\} \setminus \\
  &&  \{(t,[\rmv,e])  \in s | \\
  &&  \hphantom{s} \{ op | (t', [op, e]) \in s \land t' \neq t\} \neq
  \{ \add \} \} \setminus \\
  && \{(\bot,[\rmv,e]) \in s| \forall (t',[\add,e]) \in s \cdot t' = t \} \\ 
  \eval_i(\elems, s) &=& \{ v | (\_, [\add,v]) \in s \land
                                 (\_, [\rmv,v]) \not\in s \} \\
\end{eqnarray*}
\caption{Compact Pure \rwset CRDT}
\label{fig:rwset-compact}
\end{figure}

After invoking $\stabilize$ and removing all redundant operations, the $\stable$ function 
will replace the stable timestamps with $\bot$. Finally, the $\elems$ operation will
return all the items that are associated with $\add$ if there is no corresponding $\rmv$
in the \polog.

\section{Discussion}
\label{sec:discuss}

We recap this work by discussing the most substantial benefits the Pure 
CRDT model brings on the conceptual and performance levels. 
We divide our discussion into the following three subsections. 

\subsection{Making Op-based CRDTs Op-based}
The first incentive behind this work was to remove the confusion between
op-based and state-based CRDTs~\cite{crdt:shapiro:2011,delta-states:Almeida:2016}.
In fact, the nomenclature of these two approaches refers to the type of
the message being disseminated: the message is generally (supposed to be) the ``operation'' in
the operation-based model whereas a ``state'' in the state-based model.
This notion is, however, not reflected in the classical op-based design since
the disseminated message, i.e., the output of $\prepare$, comprises other meta-data 
as we have seen in Figure~\ref{fig:op-set}, Section~\ref{sec:back}.
In many cases, this leads to other issues (discussed next) like ad-hoc 
designs and abuse of resources.
The Pure op-based CRDTs we presented are ``pure'' op-based designs that only 
require the operation (and potential parameters) to be disseminated; thus bringing 
back the essence of the ``op-based'' terminology. The tangible benefits of this
conceptual feature is reflected in the following sections.

\subsection{Making Op-based CRDT Designs (Almost) Generic}
The Pure CRDT framework makes the design of op-based CRDTs ``almost'' generic.
In our experience, making them fully generic is impossible due to the native semantic 
discrepancy of the designed data types --- making them more generic will be 
impractical as we've seen in the \polog based CRDTs before compaction (presented in 
Section~\ref{sec:pure}). 

The classical op-based CRDT designs are data type dependent. Observing the 
seminal CRDT designs in~\cite{crdt:shapiro:2011} shows that designing a new
CRDT will require the designer/developer to completely define every design 
component, i.e., $\Sigma$, $\prepare$ for each operation type, $\effect$ 
for each operation type, and $\eval$. In the Pure CRDT framework we introduced,
the basic components are generic: $\Sigma$, $\prepare$, $\effect$, and $\stable$. 
Although our approach requires additional data-type-specific components like 
$\R$, $\R_\_$, and $\stabilize$, we argue that their definitions are very simple;
and as you may observe in the catalog of CRDTs in the previous section, these
mere components can be implemented in a couple of conditionals or assertions.
Nevertheless, our recommendation was not to use the \polog for commutative CRDTs,
being simple and more efficient, even if it will be generic.

Finally, introducing a generic framework for CRDTs brings notable benefits.
First, it reduces the complexity of the design and allows the designer/developer to 
focus on the semantics of a data type rather than fiddle with
other design details. Second,
it reduces the implementation cost since the common components can now be
implemented once and used by all data types, avoiding repetitions. Third, it reduces 
the likelihood of introducing new bugs and anomalies as the design limits the
freedom of the developer to change the rigorously thought-out generic components. 

\subsection{Making Op-based CRDT Efficient}
Our Pure CRDT model improves the efficiency of dissemination and storage.
In classical op-based CRDT designs, additional meta-data is usually used
to express the causality information required by non commutative 
data types~\cite{crdt:shapiro:2011}.
An example we've seen in this paper is the \awset, in Figure~\ref{fig:op-set},
where the state
$\Sigma = \nat \times \pow{I \times \nat \times V}$ comprises a set of
triples, and the output of $\prepare$ for the $\remove$ operation (which must be
broadcast to other replicas) contains also a set of triples. On the 
dissemination front, we believe that there is abuse of network resources since
the causality meta-data (mainly timestamps) is already being exchanged, though not used,
in the underlying middleware. The introduction of Tagged Causal Stable Broadcast
middleware (TCSB) exploits this information which allows designing pure CRDTs, i.e.,
where the $\prepare$ output (to be disseminated) is only the operation name and 
potential arguments. The dissemination overhead can be reduced several orders of
magnitude when the fraction of data to timestamps is small, e.g., when the data is
an integer whereas the timestamp is a vector clock.
TCSB has also other advantages on the storage front through
the causal stability information that helps pruning the useless meta-data in the \polog,
which as we've seen, turns the CRDT into a classical sequential data type in some cases.
As explained in Section~\ref{sec:compaction}, the gain will be higher in data types 
that exhibit large storage sizes (e.g,
as set of millions of items) since most of the operations will already be causally 
stable and thus compacted. Measuring the real overhead across real data patterns is 
subject future work.

\section{Related Work}
\label{sec:related}

\emph{Causal broadcast}
The early definitions of \emph{causality} were introduced by Lamport
in~\cite{con:rep:615}. Schiper et al. introduced an algorithm
in~\cite{Schiper:1989:NAI:645946.675010} to implement \emph{causal ordering}
based on message delivery. The authors distinguished it from Lamports
causality definitions which they called \emph{causal timestamping}. Then,
causal broadcast has been explored in the context of process group
communication in~\cite{pro:pan:160} to ensure delivery orders that do not
contradict potential causality \cite{con:rep:615}. Implementations can
piggyback to messages their causally preceding messages or tag it with a
vector clock \cite{Schmuck88theuse,alg:rep:738,alg:rep:738bis} and ensure that
preceding messages are delivered first. 

\emph{Weakly Consistent Replication.}
The design of replicated systems that are always available and eventually
converge can be traced back to historical designs
in~\cite{app:rep:optim:1501,db:rep:optim:1454,usenet:1986:NCN:6617.6618},
among others.
Lazy Replication~\cite{Ladin:1992:PHA:138873.138877} allows enforcing causal
consistency, but may apply concurrent operations in different orders in
different replicas, possibly leading to divergence if operations are not
commutative; TSAE~\cite{pan:pro:optim:813} also either applies concurrent
operations in possibly different orders, or allows enforcing a total order
compatible with causality, at the cost of delaying message delivery. Both
these systems use a message log, the former with complete causality
information, but the log is \emph{pre-delivery}, unseen by the application:
operations are applied sequentially to the current state and queries use only
the state. In our framework the \polog is \emph{post-delivery}, being part of
the data type state, maintains causality information and is used in query
operations.

\emph{Conflict-Free Replicated Data Types.}
The formalization of the commutativity requirements for concurrent operations
in replicated data types was introduced in~\cite{alg:rep:sh132,syn:app:1649},
and that of the state based semi-lattices was presented in~\cite{app:1639}.
Afterwards, the integration of the two models with many extensions was
presented in Conflict-free Replicated
Datatypes~\cite{rep:syn:sh138,syn:rep:sh143}. Currently, CRDTs have made their
way into the industry through designing highly available scalable systems in
cloud databases like RIAK~\cite{riak+crdt}, and mobile gaming industry such as
Rovio~\cite{rovio:2013}.

\emph{Message Stability.}
The notion of message stability was defined
in~\cite{Birman:1991:LCA:128738.128742} to represent a message that has been
received by all recipients; each replica can discard any message it knows to
be stable after delivering it. Similar notions are used in Lazy
Replication~\cite{Ladin:1992:PHA:138873.138877} and
TSAE~\cite{pan:pro:optim:813}. In all these cases the aim is message
garbage collection. Our definition of \emph{causal stability} is the stronger
notion that \emph{no more concurrent messages will be delivered}; we
use it inside the Datatypes to discard causality information while keeping
the operation. Causal stability is close to what is used in the mechanics of
Replicated Growable Arrays (RGA)~\cite{syn:app:1649}, although no definition
is presented there.

\emph{Message Obsolescence.}
Semantically reliable multicast~\cite{Pereira:2003:SRM:642778.642783} uses the
concept of \emph{message obsolescence} to avoid delivering messages made
redundant by some newly arrived message, where obsolescence is a strict
partial order that is a subset of causality, possibly relating messages from the
same sender or totally ordered messages from different senders. Our
obsolescence relation is more general, being defined on clock-operation pairs,
and can relate concurrent messages. Also, it is defined per-data-type, being
used inside each data type, post-delivery.

\section{Conclusions}
\label{sec:conclusion}

We introduced Pure operation-based CRDTs: a novel generic and efficient
op-based CRDT model that establishes a clear frontier with state-based models.
The model is ``pure'' in the sense that a disseminated message only contains 
the operation name and its potential arguments. The causal information that
is usually explicitly retained in the CRDT state and added to disseminated messages, 
in the classical op-based CRDT, are now provided by a  
Tagged Causal Stable Broadcast (TCSB) middleware that we proposed. Despite the useful design and
performance benefits it brings to CRDTs, we tried to introduce the TCSB in 
a self-contained section as we believe it is an important extension that 
can be exploited by the community to improve other concurrent designs as well.
We also believe that the TCSB deserves a dedicated practical analysis in the future
to understand its characteristics and performance tradeoffs.

Our paper distinguishes between two types of CRDTs: commutative and non commutative CRDTs.
For the non commutative CRDTs, we design the state as a partially ordered log (\polog)
that can retain concurrent operations tagged with timestamps. The timestamps and
some operations can be discarded using causal redundancy and stabilization that are 
features supported by the TCSB. 
On the other hand, the commutative CRDTs are not based on \polog 
being very simple. Indeed, they can be designed using the \polog if desired, we however 
don't recommend this due to the needless overhead and complexity of the \polog in this case.

We tried to provide a catalog of Pure CRDTs addressing many useful data types in
practice. We believe that these designs are now simpler, provided that the
generic framework is well understood. Some data types, like Remove-Wins Sets 
are however more complex due to their non-trivial concurrent semantics especially
when used with $\clear$ operations that reset the state. We noticed that the framework can be easier if 
reset operations are not needed. 
We did not address more composite data types like Maps since this will
require extending the \polog, probably to a Multi-\polog, thus imposing more complexity
to the paper, which we opt to avoid. Finally, we believe that a detailed 
empirical analysis that studies the different data access patterns to the CRDTs in
different models will help better understand the performance tradeoffs; we plan to 
do such analysis in the future.

\bibliographystyle{plain}
\bibliography{ref,predef,bib,shapiro-bib,local}

\begin{thebibliography}{10}

\bibitem{delta-states:Almeida:2016}
Paulo~S{'{e}}rgio Almeida, Ali Shoker, and Carlos Baquero.
\newblock Delta state replicated data types.
\newblock {\em CoRR}, abs/1603.01529, 2016.

\bibitem{causal:Attiya:2017}
H.~Attiya, F.~Ellen, and A.~Morrison.
\newblock Limitations of highly-available eventually-consistent data stores.
\newblock {\em IEEE Transactions on Parallel and Distributed Systems},
  28(1):141--155, Jan 2017.

\bibitem{Bailis:2013:ECT:2460276.2462076}
Peter Bailis and Ali Ghodsi.
\newblock Eventual consistency today: Limitations, extensions, and beyond.
\newblock {\em Queue}, 11(3):20:20--20:32, March 2013.

\bibitem{app:1639}
Carlos Baquero and Francisco Moura.
\newblock Using structural characteristics for autonomous operation.
\newblock {\em Operating Systems Review}, 33(4):90--96, 1999.

\bibitem{rep:opt:sh150}
Annette Bieniusa, Marek Zawirski, Nuno Pregui\c{c}a, Marc Shapiro, Carlos
  Baquero, Valter Balegas, and S{\'e}rgio Duarte.
\newblock Brief announcement: Semantics of eventually consistent replicated
  sets.
\newblock In Marcos~K. Aguilera, editor, {\em Int.\ Symp.\ on Dist.\ Comp.\
  (DISC)}, volume 7611 of {\em Lecture Notes in Comp.\ Sc.}, pages 441--442,
  Salvador, Bahia, Brazil, October 2012. {S}pringer-{V}erlag.

\bibitem{causal-multicast:Birman:1991}
Kenneth Birman, Andr{\'e} Schiper, and Pat Stephenson.
\newblock Lightweight causal and atomic group multicast.
\newblock {\em ACM Trans. Comput. Syst.}, 9(3):272--314, August 1991.

\bibitem{Birman:1991:LCA:128738.128742}
Kenneth Birman, Andr{\'e} Schiper, and Pat Stephenson.
\newblock Lightweight causal and atomic group multicast.
\newblock {\em ACM Trans. Comput. Syst.}, 9(3):272--314, August 1991.

\bibitem{pro:pan:160}
Kenneth~P. Birman and Thomas~A. Joseph.
\newblock Reliable communication in the presence of failures.
\newblock {\em Trans.\ on Computer Systems}, 5(1):47--76, February 1987.

\bibitem{birman:1987:reliable}
Kenneth~P Birman and Thomas~A Joseph.
\newblock Reliable communication in the presence of failures.
\newblock {\em ACM Transactions on Computer Systems (TOCS)}, 5(1):47--76, 1987.

\bibitem{DBLP:conf/popl/BurckhardtGYZ14}
Sebastian Burckhardt, Alexey Gotsman, Hongseok Yang, and Marek Zawirski.
\newblock Replicated data types: specification, verification, optimality.
\newblock In Suresh Jagannathan and Peter Sewell, editors, {\em POPL}, pages
  271--284. ACM, 2014.

\bibitem{riak+crdt}
Sean Cribbs and Russell Brown.
\newblock Data structures in {R}iak.
\newblock In {\em Riak Conference (RICON)}, San Francisco, CA, USA, oct 2012.

\bibitem{riak-crdt:Cribbs:2012}
Sean Cribbs and Russell Brown.
\newblock Data structures in {R}iak.
\newblock In {\em Riak Conference (RICON)}, San Francisco, CA, USA, oct 2012.

\bibitem{app:rep:optim:1606}
Giuseppe DeCandia, Deniz Hastorun, Madan Jampani, Gunavardhan Kakulapati,
  Avinash Lakshman, Alex Pilchin, Swaminathan Sivasubramanian, Peter Vosshall,
  and Werner Vogels.
\newblock {D}ynamo: {A}mazon's highly available key-value store.
\newblock In {\em Symp.\ on Op.\ Sys.\ Principles (SOSP)}, volume~41 of {\em
  Operating Systems Review}, pages 205--220, Stevenson, Washington, USA,
  October 2007. Assoc.\ for Computing Machinery.

\bibitem{alg:rep:738bis}
C.~J. Fidge.
\newblock Timestamps in message-passing systems that preserve the partial
  ordering.
\newblock In {\em 11th Australian Computer Science Conference}, pages 55--66,
  University of Queensland, Australia, 1988.

\bibitem{cap:Gilbert:2002}
Seth Gilbert and Nancy Lynch.
\newblock Brewer's conjecture and the feasibility of consistent, available,
  partition-tolerant web services.
\newblock {\em SIGACT News}, 33(2):51--59, 2002.

\bibitem{pan:pro:optim:813}
Richard~A. Golding.
\newblock {\em Weak-consistency group communication and membership}.
\newblock PhD thesis, University of California Santa Cruz, Santa Cruz, CA, USA,
  December 1992.
\newblock Tech. Report no.~UCSC-CRL-92-52.

\bibitem{golding1992weak}
Richard~Andrew Golding.
\newblock {\em Weak-consistency group communication and membership}.
\newblock PhD thesis, Citeseer, 1992.

\bibitem{db:rep:optim:1454}
Paul~R. Johnson and Robert~H. Thomas.
\newblock The maintenance of duplicate databases.
\newblock Internet Request for Comments RFC 677, Information Sciences
  Institute, January 1976.

\bibitem{Ladin:1992:PHA:138873.138877}
Rivka Ladin, Barbara Liskov, Liuba Shrira, and Sanjay Ghemawat.
\newblock Providing high availability using lazy replication.
\newblock {\em ACM Trans. Comput. Syst.}, 10(4):360--391, November 1992.

\bibitem{con:rep:615}
Leslie Lamport.
\newblock Time, clocks, and the ordering of events in a distributed system.
\newblock {\em Communications of the {ACM}}, 21(7):558--565, July 1978.

\bibitem{alg:rep:sh132}
Mihai Letia, Nuno Pregui\c{c}a, and Marc Shapiro.
\newblock {CRDTs}: Consistency without concurrency control.
\newblock Rapp.\ Rech. RR-6956, Institut National de la Recherche en
  Informatique et Automatique (INRIA), Rocquencourt, France, June 2009.

\bibitem{syn:rep:1677}
Prince Mahajan, Lorenzo Alvisi, and Mike Dahlin.
\newblock Consistency, availability, and convergence.
\newblock Technical Report UTCS TR-11-22, Dept.\ of Comp.\ Sc., The U.\ of
  Texas at Austin, Austin, TX, USA, 2011.

\bibitem{alg:rep:738}
Friedmann Mattern.
\newblock Virtual time and global states of distributed systems.
\newblock In {\em Int.\ W.\ on Parallel and Distributed Algorithms}, pages
  215--226. Elsevier Science Publishers B.V. (North-Holland), 1989.

\bibitem{deltas:Almeida:2015}
{Paulo Sergio Almeida, Ali Shoker, and Carlos Baquero.}
\newblock {Efficient State-based CRDTs by Delta-Mutation}.
\newblock In {\em {Proceedings of the International Conference of Networked
  sYStems}}, NETYS'15. Springer, May 2015.

\bibitem{Pereira:2003:SRM:642778.642783}
Jos{\'e} Pereira, Lu\'{\i}s Rodrigues, and Rui Oliveira.
\newblock Semantically reliable multicast: Definition, implementation, and
  performance evaluation.
\newblock {\em IEEE Trans. Comput.}, 52(2):150--165, February 2003.

\bibitem{usenet:1986:NCN:6617.6618}
John~S. Quarterman and Josiah~C. Hoskins.
\newblock Notable computer networks.
\newblock {\em Commun. ACM}, 29(10):932--971, October 1986.

\bibitem{syn:app:1649}
Hyun-Gul Roh, Myeongjae Jeon, Jin-Soo Kim, and Joonwon Lee.
\newblock {R}eplicated {A}bstract {D}ata {T}ypes: Building blocks for
  collaborative applications.
\newblock {\em Journal of Parallel and Dist. Comp.}, 71(3):354--368, March
  2011.

\bibitem{rovio:2013}
{Rovio Entertainment Ltd.}
\newblock Rovio gaming.
\newblock {http://www.rovio.com/en}, 2013.

\bibitem{Schiper:1989:NAI:645946.675010}
Andr{\'e} Schiper, Jorge Eggli, and Alain Sandoz.
\newblock A new algorithm to implement causal ordering.
\newblock In {\em Proceedings of the 3rd International Workshop on Distributed
  Algorithms}, pages 219--232, London, UK, UK, 1989. Springer-Verlag.

\bibitem{Schmuck88theuse}
Frank~Bernhard Schmuck.
\newblock The use of efficient broadcast protocols in asynchronous distributed
  systems.
\newblock Technical Report TR 88-928, Cornell University, 1988.

\bibitem{rep:syn:sh138}
Marc Shapiro, Nuno Pregui\c{c}a, Carlos Baquero, and Marek Zawirski.
\newblock A comprehensive study of {C}onvergent and {C}ommutative {R}eplicated
  {D}ata {T}ypes.
\newblock Rapp.\ Rech. 7506, Institut National de la Recherche en Informatique
  et Automatique (INRIA), Rocquencourt, France, January 2011.

\bibitem{crdt:shapiro:2011}
Marc Shapiro, Nuno Pregui\c{c}a, Carlos Baquero, and Marek Zawirski.
\newblock A comprehensive study of {C}onvergent and {C}ommutative {R}eplicated
  {D}ata {T}ypes.
\newblock Technical Report 7506, January 2011.

\bibitem{syn:rep:sh143}
Marc Shapiro, Nuno Pregui\c{c}a, Carlos Baquero, and Marek Zawirski.
\newblock Conflict-free replicated data types.
\newblock In Xavier D{\'e}fago, Franck Petit, and V.~Villain, editors, {\em
  Int.\ Symp.\ on Stabilization, Safety, and Security of Distributed Systems
  (SSS)}, volume 6976 of {\em Lecture Notes in Comp.\ Sc.}, pages 386--400,
  Grenoble, France, October 2011. {S}pringer-{V}erlag.

\bibitem{syn:optim:rep:1433}
Douglas~B. Terry, Marvin~M. Theimer, Karin Petersen, Alan~J. Demers, Mike~J.
  Spreitzer, and Carl~H. Hauser.
\newblock Managing update conflicts in {B}ayou, a weakly connected replicated
  storage system.
\newblock In {\em Symp.\ on Op.\ Sys.\ Principles (SOSP)}, pages 172--182,
  Copper Mountain, CO, USA, December 1995. ACM SIGOPS, ACM Press.

\bibitem{cassandra:2016}
{The Apache Software Foundation}.
\newblock Apache cassandra database.
\newblock {http://cassandra.apache.org/}, 2016.

\bibitem{rep:syn:pan:1624}
Werner Vogels.
\newblock Eventually consistent.
\newblock {\em {ACM} {Q}ueue}, 6(6):14--19, October 2008.

\bibitem{app:rep:optim:1501}
Gene T.~J. Wuu and Arthur~J. Bernstein.
\newblock Efficient solutions to the replicated log and dictionary problems.
\newblock In {\em Symp.\ on Principles of Dist.\ Comp.\ (PODC)}, pages
  233--242, Vancouver, BC, Canada, August 1984.

\end{thebibliography}

\newpage

\section*{Additional data types}
\label{app:a}

\subsection*{Commutative Data Types}

\subsubsection*{Grow-only Counter (\gctr)}

\begin{figure}[b]
\begin{eqnarray*}
\Sigma = \nat && \sigma_i^0 = 0 \\
\prepare_i(\inc,n ) & = & \inc \\
\effect_i(\inc, n ) & = & n+1 \\
\eval_i(\af{value},n ) & = & n\\
\end{eqnarray*}
\caption{Pure Grow-only Counter CRDT (\gctr).}
\label{fig:gctr}
\end{figure}

As its name indicates, a \gctr is a counter that can only be incremented. 
Among its use-cases are implementing vector clocks in distributed systems 
or counting the number of clients visiting a web-site.
In \gctr, the state is usually an integer $n\in \nat$ and the only mutating 
operation is $\inc$,
which reads the current state and increments it by $1$. Since 
$+$ is commutative over $\nat$, \gctr is a commutative data type.

Figure~\ref{fig:gctr} depicts a pure CRDT design of \gctr. 
The design is self explanatory: the state is an integer $n\in \nat$ 
initialized to $0$; $\prepare$ can only be $\inc$ without any meta-data 
(and thus the design is pure).
The $\effect$ function simply increments $n$ by $1$, whereas $\eval$ invokes 
the read operation $\rd$ which returns $n$.

\subsection*{Non-Commutative Data Types}

\subsubsection*{Flags (\ewfl and \dwfl)}
A flag is a datatype that retains a single boolean value indicating 
whether it is ``enabled'' or ``disabled''. Flags~\cite{riak-crdt:Cribbs:2012} have several uses in 
practice, e.g., showing if an email has been read.
Since the two allowed mutating operations, i.e., $\enable$ and $\disable$,
can be concurrent and result in opposite outputs depending on the delivery
order, a flag is obviously a non commutative CRDT. To ensure a consistent result on
all replicas when $\enable$ and $\disable$ operations are concurrent, a design
decision is usually made to break this tie,
thus yielding two flag versions:
Enable-Wins Flag (\ewfl) where two concurrent $\enable$ and $\disable$
operations leave the flag ``enabled'', to the contrary of a 
Disable-Wins Flag (\dwfl).
The formal concurrent semantics of \ewfl and \dwfl as well as their pure CRDT 
designs are depicted in Figures~\ref{fig:ewflag-semantics},~\ref{fig:ewflag-compact}, 
and~\ref{fig:dwflag-semantics},~\ref{fig:dwflag-compact}, respectively. 

The \ewfl semantics (in Figure~\ref{fig:ewflag-semantics}) says that 
a $\af{read}$ operation on the
\polog must return ``enabled'' iff there is an $\enable$ operation
that is not strictly followed by a $\disable$ or $\clear$ (i.e., a reset) 
operation in its causal future; meaning that, 
a concurrent $\disable$ or $\clear$ will not ``disable'' the flag. 
Consequently, the first clause in the pure design in 
Figure~\ref{fig:ewflag-compact} says that a newly delivered operation is always
considered redundant, and never added to the \polog, if it is $\disable$ 
or $\clear$. The second clause, however, compares
the existing operations in the \polog to remove any operation $o$ that is strictly
followed by another operation $o'$ in its causal future, i.e., if $t<t'$. 
Since only $\enable$ operations are retained by the \polog (as the first 
clause specifies), the second clause will actually remove any $\enable$ 
operations in the \polog with the $t<t'$ condition.
Consequently, the output of the $\af{read}$ operation,
invoked by $\eval$, returns an existing $\enable$ operation, or empty otherwise,
yielding the same output of the $\af{read}$ in the base-line semantics in 
Figure~\ref{fig:ewflag-semantics}. Notice that $\stabilize$ 
has no effect in this datatype, and thus always returns the \polog as is; 
which means that 
$\stable$ only replaces the stable timestamps in the \polog with $\bot$. 

\begin{figure}[t]
\begin{eqnarray*}
  \eval_i(\af{read}, s) & = & \exists (t, \enable) \in  s \cdot
  \forall (t', o')\in s \cdot \\&& (o'=\disable \lor \clear) \cdot t \not< t'
\end{eqnarray*}
\caption{\ewfl concurrent semantics}
\label{fig:ewflag-semantics}
\end{figure}
\begin{figure}[t]
\begin{eqnarray*}
	(t,o) \R s &=& o[0] = (\disable \lor \clear) \\
  (t',o') \R_\_ (t,o) &=& t' < t \\
  \stabilize_i(t,s) &=& s \\
  \eval_i(\af{read}, s) &=& (\_, \enable) \in s \\
\end{eqnarray*}
\caption{Compact \ewfl Pure CRDT}
\label{fig:ewflag-compact}
\end{figure}

\begin{figure}[H]
\begin{eqnarray*}
  \eval_i(\af{read}, s) & = & \exists (t, \enable) \in  s \cdot
  \forall (t',\disable)\in s \cdot t' < t \\ && \land
  \forall (t'',\clear)\in s \cdot t \not< t''
\end{eqnarray*}
\caption{\dwfl concurrent semantics}
\label{fig:dwflag-semantics}
\end{figure}
\begin{figure}[H]
\begin{eqnarray*}
	(t,o) \R s &=& o[0] = \clear \\
  (t',o') \R_\_ (t,o) &=& t' < t \\
  \stabilize_i(t,s) &=& s \\
  \eval_i(\af{read}, s) &=& (\_, \enable) \in s  \land (\_, \disable) \not\in s \\
\end{eqnarray*}
\caption{Compact \dwfl Pure CRDT}
\label{fig:dwflag-compact}
\end{figure}

As for \dwfl, the concurrent specs (in Figure~\ref{fig:dwflag-semantics})
says that the flag is enabled as long as there is at least one $\enable$ operation
where all existing $\disable$ operations are in its strict causal past, 
and all existing $\clear$ operations are either concurrent or in its causal past.
Consequently, the first clause in the pure CRDT design, in 
Figure~\ref{fig:dwflag-compact}, only removes newly delivered $\clear$ operations.
The second clause however specifies that $\disable$ and $\enable$ operations are
only removed from the \polog if they are succeeded by another operation with a 
greater timestamp. To reflect this behavior, the $\af{read}$ operation returns
the $\enable$ operations such that no $\disable$ operations exist in the
\polog (in which case a $\disable$ should prevail), yielding an equivalent output
to the base-line $\af{read}$ operation in Figure~\ref{fig:dwflag-semantics}. 
Finally, $\stabilize$
has no effect in this datatype as well.

\end{document}